\documentclass[twocolumn,secnumarabic,amssymb, 
aps,graphicx]{revtex4-1}
\usepackage{amssymb}
\usepackage{graphicx}
\usepackage{color}
\evensidemargin \oddsidemargin
\def\0{{\bf 0}}
\def \RC {C\!_{{\scriptscriptstyle R}}}

\def \rU {U_{{\scriptscriptstyle R}}}

\newcommand{\Ba}{\mbox{\boldmath $\alpha$}}
\newcommand{\bb}{\mbox{\boldmath $\beta$}}
\newcommand{\Be}{\mbox{\boldmath $\epsilon$}}

\newcommand{\Bp}{{\bf p}}

\newcommand{\bx}{{\bf x}}
\newcommand{\bX}{{\bf X}}
\newcommand{\bz}{{\bf z}}
\newcommand{\bu}{{\bf u}}

\newcommand{\bv}{{\bf v}}
\newcommand{\bE}{{\bf E}}
\newcommand{\bB}{{\bf B}}
\newcommand{\bA}{{\bf A}}

\def \E {{\cal E}}
\def \ZM {Z_{{\scriptscriptstyle M}}}

\newcommand{\be}{\begin{equation}}
\newcommand{\ee}{\end{equation}}
\newcommand{\bea}{\begin{eqnarray}}
\newcommand{\eea}{\end{eqnarray}}
\newcommand{\ba}{\begin{array}}
\newcommand{\ea}{\end{array}}

\begin{document}


\title{A ``slingshot'' laser-driven acceleration mechanism of plasma electrons}

\author{Gaetano Fiore$^{1,3}$, \  Sergio De Nicola$^{2,3}$
   \\    
$^{1}$ Dip. di Matematica e Applicazioni, Universit\`a di Napoli ``Federico II'',\\
Complesso Universitario  Monte Sant'Angelo, Via Cintia, 80126 Napoli, Italy\\         
$^{2}$  SPIN-CNR, Complesso  MSA, Via Cintia, 80126 Napoli, Italy\\
$^{3}$         INFN, Sez. di Napoli, Complesso  MSA,  Via Cintia, 80126 Napoli, Italy
}


\begin{abstract}
We briefly report on the  recently proposed \cite{FioFedDeA14,FioDeN15}
electron acceleration mechanism named ``slingshot effect": under suitable conditions the impact of an ultra-short and ultra-intense laser pulse against the surface of a low-density plasma is expected to cause the expulsion of a bunch of superficial electrons with high energy in the direction opposite to that of the pulse propagation; this is due to the interplay of the huge ponderomotive force, huge  longitudinal field arising from charge separation, and the finite size of the laser spot. 

\end{abstract}

\maketitle

\section{Introduction}
\noindent

Today ultra-intense laser-plasma interactions allow  extremely compact acceleration mechanisms of charged particles to relativistic regimes, with numerous and extremely important potential  applications  in nuclear medicine (cancer therapy, diagnostics), research (particle physics, inertial nuclear fusion, optycs, materials science, structural biology,...), food sterilization,  transmutation of nuclear wastes,  etc.
A prominent  mechanism for electrons 
is the  {\it Wake-Field Acceleration} (WFA) \cite{Tajima-Dawson1979}:
electrons are accelerated  ``surfing" a plasma wake wave driven by 
a short laser  or charged particle beam within a low-density plasma sample
 (or matter to be  locally completely ionized into a plasma by the beam, 
more precisely a supersonic gas jet), 
and are expelled  just after the exit of the beam out of the plasma, behind
and \textit{in the same direction as the beam} (forward expulsion). 
WFA has proved to be particularly effective since 2004 in the socalled {\it bubble} (or {\it blowout})  regime; it  can produce electron bunches of  very good collimation, small
energy spread and energies of up to hundreds of MeVs \cite{ManEtAl04,GedEtAl04,FauEtAl04} 
  or more recently even GeVs  \cite{WanEtAl13,LeeEtAl14}.

\begin{figure*}
\begin{center}
\includegraphics[
width=8cm
]{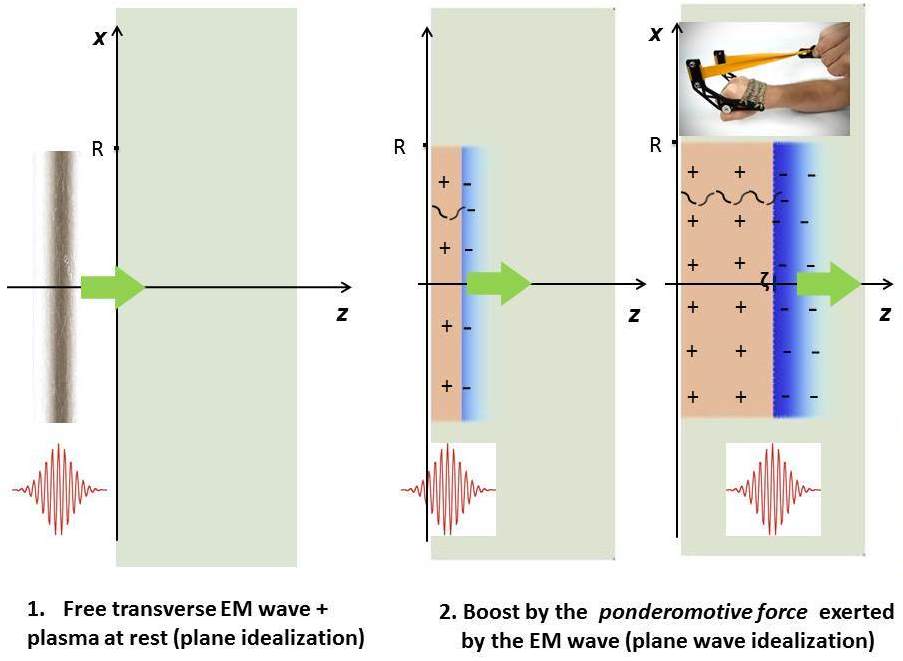} \hfill
\includegraphics[
width=8cm
]{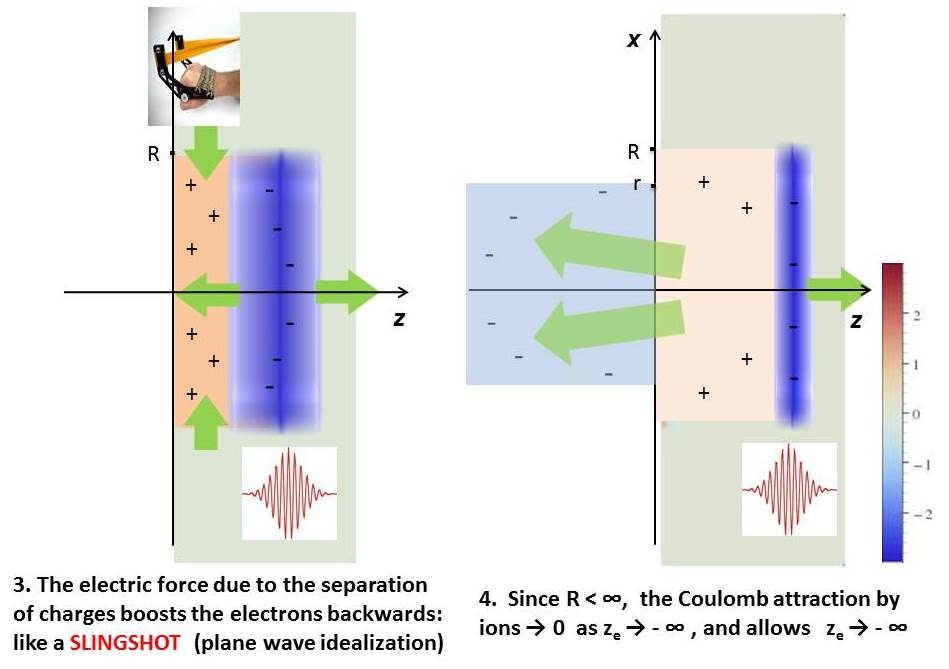}
\end{center}
\caption{Schematic stages of the slingshot effect}
\label{plasma-laser2}      
\end{figure*}

 In Ref. \cite{FioFedDeA14,FioDeN15} it has been claimed that 
the impact of a very short and intense  laser pulse 
in the form of a pancake normally onto 
the surface of a low-density  plasma may induce also the
acceleration and expulsion of electrons {\it backwards} \ ({\it slingshot effect}), see fig. \ref{plasma-laser2}.
A bunch of plasma electrons (in a thin layer just beyond the vacuum-plasma interface) 
first are displaced forward with respect to the ions  by 
the positive ponderomotive force $F\!_p\!\equiv\!\langle -e(\frac{\bv}c \times \bB)^z\rangle$
generated by the pulse (here   $\langle\: \rangle$ is 
 the average over a period of the laser carrier wave, $\bE,\bB$ are the electric and magnetic fields,
$\bv$ is the electron velocity, and  $\hat \bz$ is the direction of propagation
of the laser pulse; recall that $F\!_p$  is positive, negative when the modulating amplitude $\epsilon_s$ of the pulse  respectively increases, decreases), then  are pulled back by the electric force $-eE^z$ due to this charge displacement.
If the electron density $\widetilde{n_{0}}$ is tuned in the range where the
plasma oscillation period $T\!_{{\scriptscriptstyle H}}$ is about twice the pulse duration $\tau$, 
then these electrons invert their motion when they are reached by the maximum of $\epsilon_s$, so that the negative part of $F\!_p$ adds to  $-eE^z$ in  accelerating them backwards; equivalently, the total work 
$W\!\equiv\!\int_0^\tau \! dt \, F\!_p v^z$ done by the ponderomotive force 
is maximal\footnote{Whereas  $F\!_p v^z$ oscillates many times about 0, and $W\!\simeq\!0$, 
if \  $\tau\!\gg\! T\!_{{\scriptscriptstyle H}}$ - the standard experimental situation
until a couple of decades ago.}.
Their expulsion energy (out of the bulk) will be enough to escape to  $z\!\to\!-\infty$
if the laser spot radius $R$ is  suitably tuned.

\begin{figure}
\includegraphics[width=7cm]{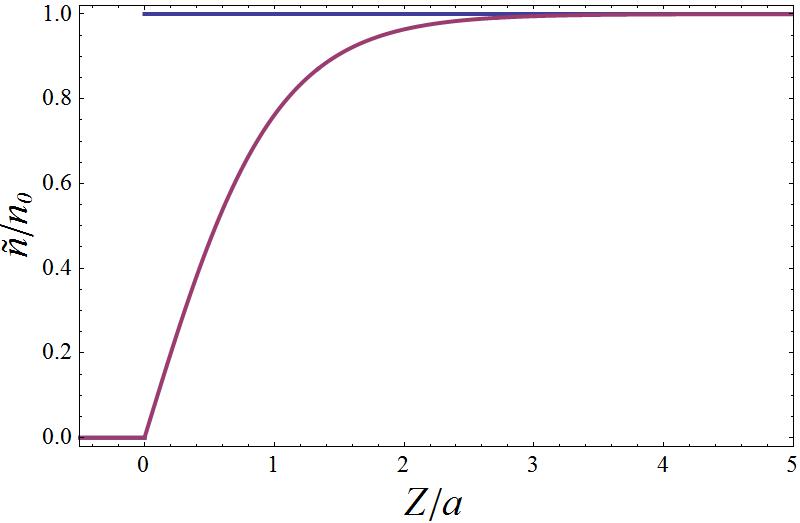} 
\caption{The normalized $\widetilde{n_{0}}$ adopted here: step-shaped  (blue) and 
continuous  $\widetilde{n_{0}}(Z)\!=\!n_0\,\theta(Z)\tanh(Z/a)$,   $a\!=\!20\mu$m (purple); they 
respectively model the initial electron  densities at the vacuum interfaces of an aerogel and of a gas jet
(just outside the nozzle).}
\label{densities}
\end{figure}

The very short pulse duration $\tau$ and expulsion time $t_e$, as well as huge nonlinearities,
make   approximation schemes  based on  Fourier analysis and related methods
unconvenient. But recourse to full kinetic theory  is not necessary: we show \cite{Fio16,FioDeN15} that
 in the relevant space-time region
a MagnetoHydroDynamic (MHD) description of the impact  is self-consistent, simple and predictive. 
The set-up is as follows.
We describe the plasma as consisting of a static background of ions 
and a fully relativistic, collisionless fluid of electrons, with the system ``plasma + electromagnetic field"  fulfilling 
the Lorentz-Maxwell and the continuity Partial Differential Equations (PDE). 
For brevity, below we  refer to the electrons'
fluid element initially located at $\bX\!\equiv\!(X,Y,Z)$ as to the  ``$\bX$ electrons", and to the fluid elements with arbitrary $X,Y$ and specified $Z$  as the ``$Z$ electrons".
We denote: as \ $\bx_e(t,\bX)$  \ the position at time $t$
of the \ $\bX$ electrons, \ and for each fixed $t$ as
$\bX_e(t,\bx)$ the inverse of  $\bx_e(t,\bX)$  [$\bx\!\equiv\!(x,y,z)$];
as $c$ the velocity of light; as $m$ and
as $n,\bv,\Bp$ the electrons' mass and Eulerian
density, velocity, momentum.  \ $\bb\!\equiv\!\bv/c$,
$\bu\!\equiv\!\Bp/mc\!=\!\bb/\sqrt{1\!-\!\bb{}^2}$, 
$\gamma\!\equiv\!1/\sqrt{1\!-\!\bb{}^2}\!=\!\sqrt{1\!+\!\bu^2}$ \ are dimensionless.
We assume that the plasma is initially 
neutral, unmagnetized and at rest with electron (and proton) density $\widetilde{n_{0}}(z)$
depending only on $z$ and
equal to zero  in the region  \ $z\!<\! 0$. 
We schematize the laser pulse  as a free transverse EM plane travelling-wave
 multiplied by a cylindrically symmetric ``cutoff'' function, e.g.
\be
\bE^{{\scriptscriptstyle\perp}}\!(t,\!\bx)=\Be\!^{{\scriptscriptstyle\perp}}\!(ct\!-\!z)\,
\theta(R\!-\!\rho),\qquad  \bB^{{\scriptscriptstyle\perp}}=
{\hat{\bf z}}\!\times\!\bE\!^{{\scriptscriptstyle\perp}}                                                  \label{pump}
\ee
where  $\rho\!\equiv\sqrt{\!x^2\!+\!y^2}\!\le\! R$, 
$\theta$ is the Heaviside step function, and
the  `pump' function  $\Be^{{\scriptscriptstyle\perp}}(\xi)$ \ vanishes
outside some finite interval \ $0\!<\!\xi\!<\!l$.
Then, to simplify the problem,

\begin{enumerate}

\item We first study  the $R\!=\!\infty$ (i.e. {\it plane-symmetric}) version,
carefully choosing  unknowns and  independent variables (section \ref{Planewavessetup}).
For sufficiently small densities and short times 
 we can reduce  the PDE's to a  collection of decoupled {\it systems of two first order 
 nonlinear ODE in  Hamiltonian form}, which we  solve numerically.

\item  We determine (section \ref{3D-effects}): 
$R\!<\!\infty$, $r\!>\!0$ so that the plane version   gives small errors for the surface electrons
with $\rho\!\le\!r\!\le\! R$;
 the corresponding final energy, spectrum, etc. of the expelled electrons. 
For definiteness, we consider the $\widetilde{n_{0}}(z)$ of fig. \ref{densities}.

\end{enumerate}

We specialize our  predictions to virtual experiments at the FLAME facility (LNF, Frascati).
We invite for simulations (PIC, etc.) and experiments testing them.

\section{The model}
\subsection{Plane wave idealization}
\label{Planewavessetup}

\noindent
Our plane wave Ansatz reads: \ $A^\mu, \bu,n\!-\!\widetilde{n_{0}}(z)$ \ depend only on  $z,t$
and vanish if \  $ct\!\le\! z$; \ 
$\Delta \bx_e\!\equiv\! \bx_e\!-\!\bX$ \  depends only on $Z,t$ and vanishes 
 if \  $ct\!\le\! Z$. Then:   $\bB\!=\!\bB^{{\scriptscriptstyle\perp}}\!=\!\hat\bz\partial_z\!\wedge\!\bA\!^{{\scriptscriptstyle\perp}}\!,$ \    
 $c\bE^{{\scriptscriptstyle\perp}}\!\!=\!-\partial_t\bA\!^{{\scriptscriptstyle\perp}}\!$; \  the transverse component of the Lorentz equation implies
$\bu^{{\scriptscriptstyle\perp}}\!= \!e\bA\!^{{\scriptscriptstyle\perp}}\!/m{}c^2$; \ 
due to charge separation $E^z\!\neq\!0$: by the Maxwell
equations it is related to the longitudinal motion  through
\be
E^z(t,\! z)\!=\!4\pi e \!\left\{
\widetilde{N}(z)\!-\! \widetilde{N}[Z_e (t,\! z)] \right\}, \:\: 
\widetilde{N}(Z)\!\equiv\!\!\!\int^Z_0\!\!\!\!\!\! d\eta\,\widetilde{n_{0}}(\eta),
 \label{elFL}{}
\ee
what yields a conservative  force on the electrons.
For sufficiently small densities and short times the laser pulse is 
not significantly affected by the interaction with the plasma 
(the validity of this approximation is checked a posteriori \cite{FioDeN15}),
and we can identify  \ $\bA\!^{{\scriptscriptstyle\perp}}(t,z)\!=\!\Ba(\xi)$,
 $\xi\!\equiv\! ct\!-\!z$, \ where $\Ba$ is the
transverse vector potential of the ``pump" free laser pulse.
Hence also \ $\bu\!^{{\scriptscriptstyle\perp}}(t,z)\!=\!e\Ba(\xi)/m{}c^2$ \ is explicitly determined.
For each fixed $Z$, the unknown 
$z_e(t,Z)$ appears in place of $z$  in the equations of motion of the $Z$-electrons. But, as no particle can reach the speed of light,  the map \ $t\!\mapsto\! \xi\!\equiv\! ct\!-\!z_e(t,Z)$ \ is strictly increasing, and we can  
use  \cite{FioDeN15,Fio16} {\it $(\xi,Z)$  instead of} \
$(t,\!Z)$ as independent variables.  
It is also convenient to use the \ ``electron $s$-factor''  $s\!\equiv\!\gamma\!-\! u^z$ \
instead of \ $u^z$ \ as an unknown, because it is 
{\it insensitive} to rapid oscillations of $\Ba$, \ and 
$\gamma,\bu,\bb$ are {\it rational}  functions of $\bu^{{\scriptscriptstyle\perp}},s$:
\be
\gamma\!=\!\frac {1\!+\!\bu^{{\scriptscriptstyle\perp}}{}^2\!\!+\!s^2}{2s}, 
\qquad u^z\!=\!\frac {1\!+\!\bu^{{\scriptscriptstyle\perp}}{}^2\!\!-\!s^2}{2s}, 
 \qquad \bb\!=\! \frac{\bu}{\gamma} \label{u_es_e}.
\ee
Then the remaining PDE to be solved are reduced to the following collection
of systems  (parametrized by $Z$) of  first order ODE's in the unknowns
$\Delta(\xi,Z),s(\xi,Z)$:
\bea
\!\!\! \Delta' = \displaystyle\frac {1\!+\!v}{2 s^2}\!-\!\frac 12,
\qquad 
 s'=\frac{4\pi e^2}{mc^2}\left\{
\widetilde{N}[\Delta\!+\!Z] \!-\! \widetilde{N}(Z)\right\}\label{heq1} \\[8pt]
 \Delta(0,\!Z)=0,  \qquad\quad\:
  s(0,\!Z)= 1.  \qquad\qquad\qquad \label{heq2}
\eea
Here \ $v(\xi )\!\equiv\![e\Ba(\xi)/mc^2]^2$, \
$\Delta\!\equiv\!z_e\!-\!Z$, 
$f'\!=\!\partial f/\partial \xi$.
Eq.s (\ref{heq1}) can be written also in the form \cite{Fio16} of  {\it Hamilton equations}  
\ $q'=\partial H/\partial p$, $p'=-\partial H/\partial q$ \ in 1 degree of freedom:
  $\xi,-\Delta, s$  play the role of $t,q,p$.
Solving (\ref{heq1}-\ref{heq2})  numerically
all unknowns are determined. 
For $z\!>\!0$ \ $\bu(t,z),n(t,z),...$  evolve as forward travelling waves.
\begin{figure}
\includegraphics[width=8cm]{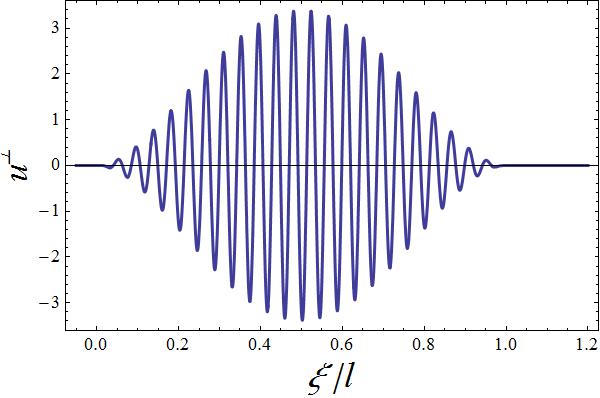} \\
\includegraphics[width=8cm]{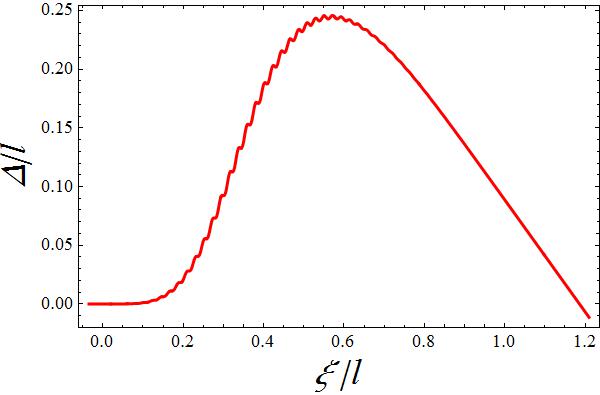} \\
.\quad\includegraphics[width=7.5cm]{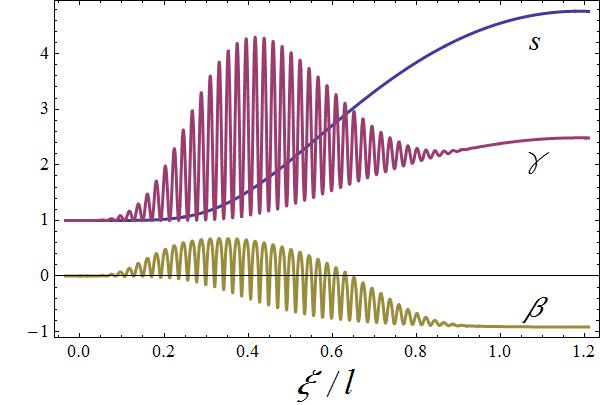} \\
\includegraphics[width=7cm]{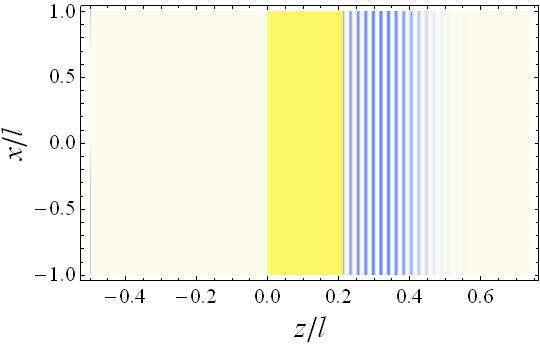}\hfill
\includegraphics[width=0.6cm]{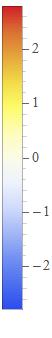} 
\caption{Typical normalized pump amplitude \ $\bu^{{\scriptscriptstyle\perp}}\!=\!e\Ba/mc^2$  
vanishing outside $0\!<\!\xi \!<\!l$ (up),  corresponding solution of (\ref{e1}-\ref{e2})  for   
$Ml^2\!=\!26$ (center) and normalized charge density plot after 40 fs (down).
}
\label{graphsb}
\end{figure}

In particular, if $\widetilde{n_{0}}(\!Z\!)\!=\!n_0\theta(\!Z\!)$  then  (\ref{elFL})
implies  that the  longitudinal electric force acting on the $Z$-electrons is
\be
\!\widetilde{F}_e^{{\scriptscriptstyle z}}(t,Z)
\!=\! \left\{\!\!\!\ba{ll}
- 4\pi n_0 e^2\!  \Delta z_e  \mbox{= elastic force}\: &\mbox{if }z_e\!>\!0,\\
 \: 4\pi n_0 e^2 Z\: \mbox{= constant force}\: &\mbox{if }z_e\!\le\!0;
\ea\right.\ee
hence   as long as $z_e\ge 0$ each $Z$-layer of electrons  is an independent copy of the {\it same} 
relativistic harmonic oscillator,  (\ref{heq1}-\ref{heq2}) are $Z$-independent and  reduce 
to a {\it single} system of two  first order ODE's
\bea
&& \Delta'=\displaystyle\frac {1\!+\!v}{2s^2}\!-\!\frac 12,\qquad\qquad
 s'=M\Delta,\label{e1}\\[6pt]
&&  \Delta(0)\!=\!0, \qquad\qquad\qquad\:\:   s(0)\!=\! 1,\label{e2}
\eea
($ M \!\equiv\!4\pi n_0e^2\!/mc^2$). \ $n_0\!\!\to\!\!0$ implies $s\!\!\equiv\!\!1$, and the equations are solved in closed form \cite{Fio14JPA,Fio14}. In fig. \ref{graphsb} we plot a typical pump and the corresponding solution of (\ref{e1}-\ref{e2}).

\subsection{Finite $R$ corrections and experimental predictions}
\label{3D-effects}

\begin{figure}
\includegraphics[width=7.5cm]{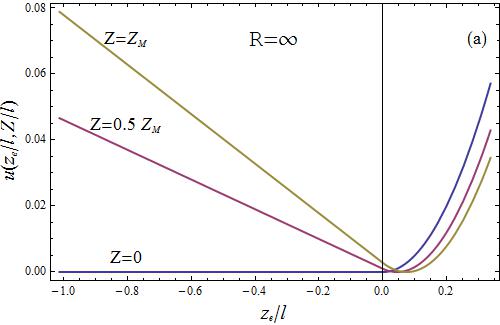} 
\includegraphics[width=7.5cm]{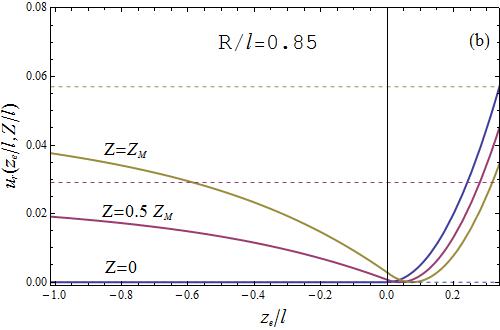}
\caption{Rescaled longitudinal electric potential energies $u\!\equiv\! U/4\pi n_0e^2\!l^2$,
$u_r\!\equiv\! \rU /4\pi n_0e^2\!l^2$ 
in the case of step-shaped density $\widetilde{n_{0}}(\!Z\!)\!=\!n_0\theta(\!Z\!)$ and
(a) idealized plane wave or (b) for $R/l\!=\!0.85$, vs. the dimensionless ratio
$z_e/l$ for
$Z\!=\!0$, $Z\!=\!0.5\ZM$,  $Z\!=\!\ZM$; the horizontal dashed lines are the left asymptotes of $u_r$ for
the same values of $Z$.}
\label{Upotentials}
\end{figure}
The results of the previous section can be applied to the small $Z$ Forward Boosted Electrons (FBE)
 in the region $\rho\!\lesssim\! R$; this leads to a cylinder $\RC$ of the same radius
completely deprived of electrons, with maximum height $\zeta$ at the 
time $\bar t$ of maximal penetration $\zeta$ of the FBE.
The displaced charges modify $\bE$. 
By causality, for $\bx$ near the $\vec{z}$ axis $\bE(t,\bx)$ 
 is the same as in  the plane wave case for $t\!\lesssim\! \bar t\!+\!R/c$ (the ``information about 
the finite $R$" contained in the retarded fields takes a time $R/c$ to go from
$\rho\!=\!R$ to $\rho\!=\!0$), and smaller afterwards. 
We tune $R$ to fulfill
\be
 \frac{[t_e \!-\!\bar t]c}R \sim 1,   \qquad
r \equiv R-\frac{\zeta(t_e\!-\!l/c)}{2(t_e\!-\!\bar t)}\: \theta(ct_e\!-\!l ) >0                                   \label{req}
\ee
($t_e$ is the expulsion time  of the FBE). Conditions (\ref{req}) respectively ensure:  that
the motion of these FBE is close to the one in section \ref{Planewavessetup} until their expulsion, 
at least  within an inner cylinder $\rho\!\le\! r\!\le\! R$; that  their expulsion takes place before
lateral electrons,
which are initially located outside the surface of 
$\RC$ and are attracted towards the $\vec{z}$-axis, obstruct them the way out 
\cite{FioFedDeA14,FioDeN15}.

\begin{figure*}
\includegraphics[width=15cm]{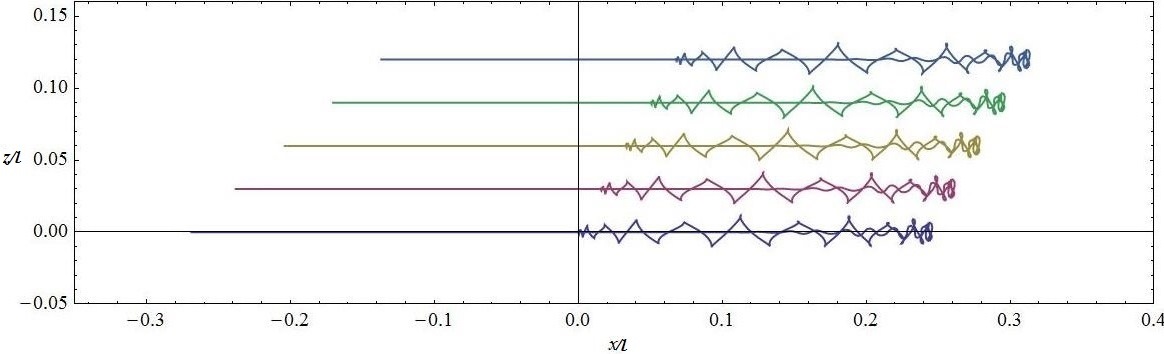} \\
\caption{Trajectories  (after about 150 fs) of electrons initially located at  different longitudinal positions: 
\  $Z/\ZM=0,  0.25 , 0.5 ,  0.75 ,1$, from bottom up.  The conditions are the same as   in fig. \ref{graphsb}.}
\label{Traj2}
\end{figure*}

The expelled $Z$ electrons are decelerated by the electric force $\widetilde{F_e^{{\scriptscriptstyle zr}}}\!>\!0$ generated by
the net positive charge located at $z\!>\!z_e(t,Z)$, but $\widetilde{F_e^{{\scriptscriptstyle zr}}}\!\propto \!1/z_e^2$ as 
$z_e \!\to\! -\infty$ since this charge is localized in $\RC$. 
Therefore we heuristically modify at $z_e \!<\! 0$ the  potential energies $U(z_e,Z)$ 
associated to (\ref{elFL}).
If e.g.  $\widetilde{n_{0}}(\!Z\!)\!=\!n_0\theta(\!Z\!)$ then 
$U(z_e,Z)\!=\!2\pi n_0e^2[ \theta(z_e)z_e^2\!-\!2z_eZ\!+\!Z^2]$
is modified for $z_e \!<\! 0$ into
$$\ba{ll}
\rU(z_e,\!Z) \!=\! & \pi n_0e^2 \!\left[(z_e\!-\!2Z)\sqrt{\! (z_e\!-\!2Z)^2\!+\!R^2}\!-\!4Zz_e \right. \\[6pt]
 &-\!z_e\sqrt{\! z_e^2\!+\!R^2}\!+\! 2Z^2  \!\!+\!2Z\sqrt{\! 4Z^2\!+\!R^2} \\[6pt]
&\left. \!+\!R^2\!\left(\!\sinh^{\scriptscriptstyle{-1}}\!\!\frac{2Z}R\!+\!\sinh^{\scriptscriptstyle{-1}}\!\!\frac{z_e\!-\!2Z}R 
 \!-\!\sinh^{\scriptscriptstyle{-1}}\!\!\frac{z_e}R\!\right)\!\right].
\ea $$
(see fig. \ref{Upotentials}).
Solving the equations of motion we find that for sufficiently small $Z$ ($0\!\le\!Z\!\le\!Z_{{\scriptscriptstyle M}}$)
the map \  $\bX\!\mapsto\! \bx_e(t,\bX)$ \  is  one-to-one for all $t$ (showing the {\it self-consistency} of this MHD treatment)
 and that $z_e(t,\!Z)\!\stackrel{t\to\infty}{\longrightarrow}\! -\infty$
(backward escape); some typical electron trajectories are shown in fig.'s
\ref{Traj2}, the animated versions are available at the hyperlink people.na.infn.it/$\sim$gfiore/slingshot-videos.
The interplay of the ponderomotive, electric forces
yield the longitudinal  forward and backward drifts at the basis of the slingshot effect.
On the contrary, transverse oscillations due to $\bE^{{\scriptscriptstyle \perp}}$ average to zero to yield  vanishing final transverse drift and momentum,
if - as usual - the pump (\ref{pump}) has a slow modulation $\epsilon_s$ in the support $0\!<\!\xi\!<\! l$: 
\be
\Be\!^{{\scriptscriptstyle\perp}}\!(\xi)
\!=\!{\hat\bx}\epsilon_s(\xi)\cos k\xi\qquad
\mbox{with }\: |\epsilon_s'|\!\ll\! |k\epsilon_s|
\label{modulation}
\ee
(here the pump is polarized  e.g.  in the $x$-direction) 
implies \   $p^{{\scriptscriptstyle \perp}}(\xi)\!\simeq\!
\epsilon_s(\xi) \left|\sin(k\xi) e/kc\right|\!=\!
0$ \ for $\xi\!\ge\! l$, and hence a good collimation of the expelled electrons.  
If the plasma is created by the impact on a supersonic gas jet (e.g. helium)  of the pulse itself, then
 $l\!<\!\infty$ is the length of the interval where the intensity is sufficient to 
ionize the gas. 

On the contrary, deeper electrons ($Z\!>\!Z_{{\scriptscriptstyle M}}$) are captured by
the force $\widetilde{F_e^{{\scriptscriptstyle zr}}}$ and circulate forth and back.

The EM energy $\E$ carried by a pulse (\ref{pump}), (\ref{modulation})  is 
\be
\E\!\simeq\!\frac{R^2}{8}\!\! \int_0^l\!\!\!\!d\xi\,\epsilon_s^2(\xi).
\label{pulseEn}
\ee
$\E$ is fixed and depends on the laser; reducing $R$ (focalization)  increases the intensity $I$, the electron 
penetration $\zeta$ and the slingshot force, but we must not violate (\ref{req}).
In fact,  (\ref{req}) can be fulfilled  with a rather small $R$. 

\begin{figure}
\includegraphics[width=7cm]{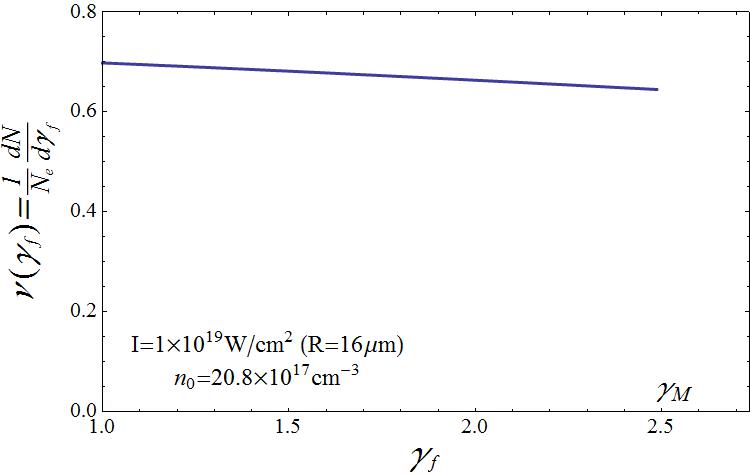} \\
\includegraphics[width=7cm]{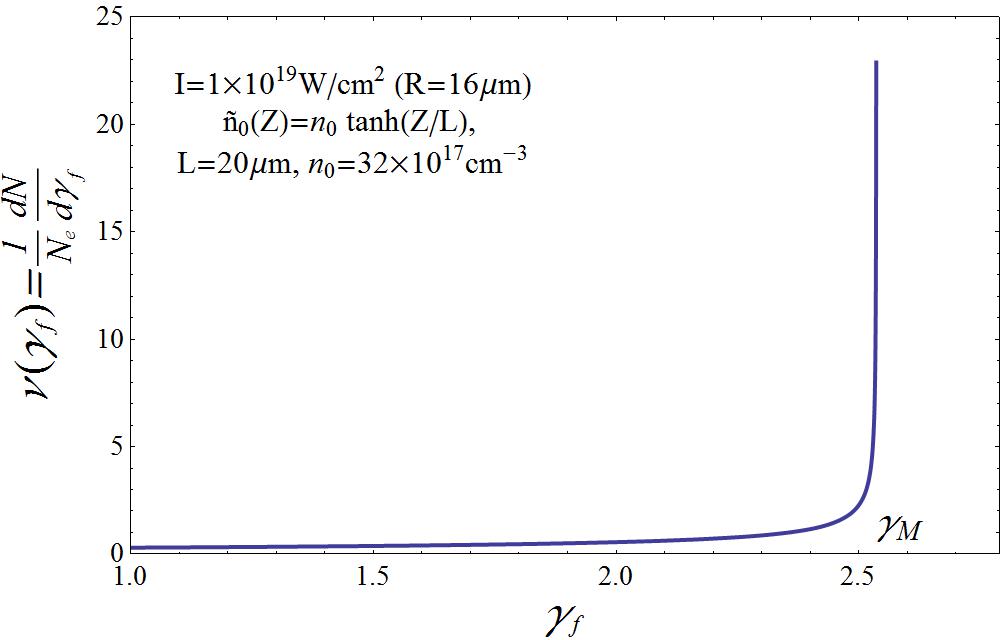} 
\caption{Spectra of the  expelled electrons for mean pulse intensitiy $I\!=$10$^{19}$ W/cm$^2$ and step-shaped
(up) or continuous (down)  $\widetilde{n_{0}}$. 
}
\label{nu_vs_gamma}
\end{figure}

We adopt a modulating amplitude of gaussian form 
\be
\epsilon_s^g(\xi)\!=\!b_g\, \exp\!\left[-(\xi\!-\!l/2)^2/2\sigma\right]\theta(\xi)\theta(l\!-\!\xi);
\ee
the parameters $b_g,\sigma,l,...$ are determined by $\E,R$ and the full width at half maximum $l'$
of the pulse.
We report in table \ref{tab1g}  and fig.  \ref{nu_vs_gamma} sample results of extensive numerical simulations
performed using as inputs the parameters available \cite{GizEtAl13} in virtual experiments at the FLAME facility 
of the INFN Laboratori Nazionali di Frascati:
\ $l'\!\simeq\! 7.5 \mu$m (corresponding to a time $\tau'\!=\!25$fs), \ wavelength $\lambda\!\simeq\! 0.8 \mu$m, 
 \ $\E\!=\!5 $J, \  
 $R$ tunable in the range \ $10^{-4}\!\div\! 1$cm; a supersonic helium jet or an aerogel  (if $\widetilde{n_{0}}(Z)\!=\!n_0\,\theta(Z)$ with $n_0\!\gtrsim\! 48\!\times\!10^{18}$cm$^{-3}$) as targets. The energy spectrum,  or equivalently the distribution $\nu(\gamma_f)$  of   the expelled electrons as a function of  their final relativistic factor, depends dramatically on $\widetilde{n_0},R$; pleasantly, in the case $\widetilde{n_{0}}(Z)\!=\!n_0\,\theta(Z)\tanh(Z/a)$ it is very peaked (almost monochromatic) around $\gamma_{{\scriptscriptstyle M}}$, the maximal $\gamma_f$.

Summarizing, this new  laser-induced ``slingshot"  acceleration mechanism should yield well-collimated bunches 
of electrons of energies up to few tens MeV. It is easily tunable  and testable with present 
equipments.

\noindent 
{\bf Acknowledgments}  
We thank  R. Fedele for stimulating discussions,
Compagnia di San Paolo for support 
 under grant {\it Star Program 2013}.

\begin{table}
\begin{tabular}{|c|}
\hline
$\!$pulse energy ${\cal E}\!\simeq\!5$J, wavelength $\lambda\!\simeq\!.8\mu$m, 
duration  $\tau'\!\!=\!25$fs$\!$ \\
\hline
\begin{tabular}{|l|c|c|c|c|c|c|c|c|c|c|}
\hline
 type of target &  h & h  & h & h & h & a &a \\[2pt]
 pulse spot radius $R \, (\mu$m)   &   16 & 8 & 4 &  2 &  2  &   2 & 1\\[2pt]
mean intensity $I$(10$^{19}$W$\!$/cm$^2$)  &  1&  4 &16 & 64&  64 & 64 & 255\\[2pt]
initial  density $n_0(10^{19}$cm$^{-3}$)  &  0.8 & 2  &  13  &    80 &   
 20 &  12 & 40\\[2pt]
max. relativistic factor $\gamma_{{\scriptscriptstyle M}}$  &  2.6 &  6&  8.5  &  
14 &   21 &  12  &  23 \\[2pt]
max. expulsion energy(MeV)  & 1.3 & 3 & 4  &   7 &  11 &  6.4 & 12 \\[2pt]
\hline
\end{tabular}
\end{tabular}
\caption {Sample inputs and corresponding outputs  if the target is: a supersonic helium jet (h)
 or an aerogel (a) with initial density profiles as in fig. \ref{densities}. The expelled charge
is in all cases a few $10^{-10}$C 
}
 \label{tab1g}  
\end{table}

\end{document}